# Negative magnetostrictive magnetoelectric coupling of BiFeO$_3$

Sanghyun Lee[1,2,3], M. T. Fernandez-Diaz[4], H. Kimura[5], Y. Noda[5], D. T. Adroja[6], Seongsu Lee[7], Junghwan Park[2,3], V. Kiryukhin[8], S-W. Cheong[8], M. Mostovoy[9], and Je-Geun Park[1,2,10,#]

[1] Center for Functional Interfaces of Correlated Electron Systems, Institute for Basic Science, Seoul National University, Seoul 151-747, Korea
[2] Center for Strongly Correlated Materials Research, Seoul National University, Seoul 151-742, Korea
[3] Department of Physics, SungKyunKwan University, Suwon 440-746, Korea
[4] Institut Laue-Langevin, B.P. 156, 38042 Grenoble Cedex, France
[5] Institute of Multidisciplinary Research for Advanced Materials, Tohoku University, Sendai 980-8577, Japan
[6] ISIS Facility, Rutherford Appleton Laboratory, Chilton, Didcot, Oxon OX11 0QX, United Kingdom
[7] Neutron Science Division, Korea Atomic Energy Research Institute, Daejeon 305-353, Korea
[8] Rutgers Center for Emergent Materials and Department of Physics and Astronomy, Rutgers University, Piscataway NJ 08854, USA
[9] Zernike Institute for Advanced Materials, University of Groningen, Nijenborgh 4, 9747 AG, Groningen, The Netherlands
[10] FPRD, Department of Physics & Astronomy, Seoul National University, Seoul 151-747, Korea

PACS: 75.85.+t, 61.05.F-, 75.50.Ee

How the magnetoelectric coupling actually occurs on a microscopic level in multiferroic BiFeO$_3$ is not well known. By using the high-resolution single crystal neutron diffraction techniques, we have determined the electric polarization of each individual elements of BiFeO$_3$, and concluded that the magnetostrictive coupling suppresses the electric polarization at the Fe site below $T_N$. This negative magnetoelectric coupling appears to outweigh the spin current contributions arising from the cycloid spin structure, which should produce a positive magnetoelectric coupling.

Multiferroic materials with the coexistence of both ferroelectricity and magnetism offer huge potential and, at the same time, pose new challenges for our understanding of how magnetism and ferroelectricity can be coupled to one another in real materials [1,2]. Once fully achieved, this understanding of their coupling, the so-called magnetoelectric effects, can lead to the better manipulation of the unusual multiferroic behavior. However, one needs to carry out precise measurements of both structure and dynamics in order to gain the fundamental understanding of the underlying physics. Despite the academic and technological importance, however the origin of the magnetoelectric coupling has proven often challenging to address experimentally for a given material.

$BiFeO_3$ is arguably one of the most extensively studied multiferroic materials with several distinctive properties [3-5]. For example, it has both magnetic and ferroelectric phase transitions above room temperature: $T_N$ = 650 K and $T_C$ = 1100 K. Moreover, it has one of the largest ever reported values of polarization, ~ 86 $\mu C/cm^2$, with the non-centrosymmetric space group of *R3c* in the ferroelectric phase as shown in Fig. 1a. However, how the magnetoelectric coupling actually occurs and moreover how to understand it remains to be resolved. Another interesting point to be noted is that when it undergoes basically a G-type magnetic ordering at 650 K an incommensurate structure is formed with an extremely long period of 620 Å due to Dzyaloshinski-Moriya (DM) interaction without breaking the crystal symmetry (see Fig.1b & 1c) [6]. In the hexagonal notation, the propagation vector of the incommensurate structure is **Q** = [0.0045, 0.0045, 0] at room temperature with a chiral vector of $\mathbf{e_3}$=[-1 1 0]. The spin waves of $BiFeO_3$ measured by inelastic neutron scattering techniques [7,8] are consistent with a Heisenberg Hamiltonian with DM interaction, which was derived from bulk measurements [9].

As regards the magnetoelectric coupling, multiferroic materials with chiral magnetic structures offer an interesting, so far unexplored, possibility of the inverse DM effects. This inverse DM effect induces an electric polarization when a particular chiral structure sets in [10]. The underlying mechanism of such additional polarization has been theoretically investigated by several groups [11,12]. Another equally viable scenario is the more classical mechanism of a lattice-mediated spin-lattice coupling. Few systematic studies have so far been made of which one of the two mechanisms works for $BiFeO_3$.

Here, we present the detailed experimental measurements of the magnetoelectric coupling in $BiFeO_3$ by using the high-resolution single crystal neutron diffraction techniques. Through the close examination of the temperature dependent electric polarizations by Bi and Fe atoms, we could unequivocally demonstrate that the magnetostrictive magnetoelectric coupling suppresses the polarization of Fe in the magnetically ordered phase, outweighing that due to the inverse DM effect.

High-resolution single crystal neutron diffraction experiments were carried out using a neutron wavelength of 0.835 Å from 300 to 850 K at the D9 diffractometer of the ILL, France. We used single crystals grown by a flux method with the typical size of 1.6 x 2.6 x 2.4 $mm^3$. We have carried out all our analysis using the Fullprof program [13].

$BiFeO_3$ of the R3c space group undergoes an incommensurate antiferromagnetic transition at $T_N$=650 K with the cycloid magnetic structure as shown in Fig. 1a. Because of the three-fold rotational symmetry along the c-axis, there are three equivalent propagation vectors for the cycloid magnetic structure with $Q_1=[1\ 1\ 0]$, $Q_2=[-2\ 1\ 0]$, and $Q_3=[1\ -2\ 0]$, and they are thermally populated as separate magnetic domains as seen by polarized neutron diffraction experiments [14]. There are two possible directions of the spin rotation for each of the three Q vectors, which are related to the DM interaction aforementioned. For example, the spin chiral vector $e_1$ parallel to the [-1 1 0] axis produces a clockwise rotation of the cycloid structure while $e_2$ parallel to the [1 -1 0] axis does an anticlockwise rotation, when the cycloid structure is viewed on the (h h l) plane. A recent x-ray scattering experiment with polarization analysis shows that the sense of the spin rotation is clockwise [15].

This clockwise spin rotation can then give rise to an induced electric polarization $\triangle P$ via the inverse DM effect. This additional electric polarization can be written in the following form: $\Delta P = Ae \times Q$, where A is a material specific coefficient, e is the spin-rotation chiral vector and Q is the propagation vector of the chiral magnetic structure [12]. Furthermore, it is important to note that a Ginzburg-Landau analysis based on the symmetry of $BiFeO_3$ predicts that $\triangle P$ should be parallel to the total polarization direction in the paramagnetic phase, i.e. positive $\triangle P$. In our calculation, the total electric polarization (P) always points along the c-axis above $T_N$. Notice that this dependence of $\triangle P$ due to the inverse DM effect is equally valid for the two other Q vectors. On the other hand, the magnetoelectric coupling of magnetostrictive nature can produce negative $\triangle P$. Therefore, one can precisely determine which one of the two mechanisms is at work for $BiFeO_3$ simply by measuring $\triangle P$ below $T_N$.

This realization opens up a simple, yet elegant way of addressing the issue of a magnetoelectric coupling in $BiFeO_3$. In fact, we have tried to answer this question by carrying out high-resolution structure studies before, where we noted a quite significant change in the lattice constants below $T_N$ [16]. However, because of technical limitations in our previous diffraction experiments, we had to make most of our analysis then using synchrotron data. More specifically, we could not examine the temperature-dependent contributions to the electric polarization by individual elements because of the lack of the accurate information of the O positions over the entire temperature range: x-ray data are typically less sensitive to lighter elements like O compared with neutrons.

In order to overcome these previous technical difficulties, we have now carried out higher-resolution single crystal neutron diffraction experiments using the D9 beamline. We have taken special efforts to increase the number of the nuclear peaks by a factor of 2 compared with the previous experiment as well as to measure the data over a very wide temperature range, in particular extending the temperature range above $T_N$.

Let us explain the aim of the experiment from the crystallographic viewpoint first. To estimate the individual electric polarization by Bi and Fe atoms, we have assumed the high temperature paraelectric phase of Pm-3m space group. Although the ferroelectric transition and the space group of the paraelectric phase are still under debate [17,18]: one of the proposed candidates is the Pbnm orthorhombic structure (see Fig. 2), our analysis and conclusion below are valid regardless of the nature of the ferroelectric transition since we only concern here with the temperature-induced electric polarization. With the paraelectric phase of Pm-3m, the electric polarization by individual atoms can be calculated by simply measuring the relative shifts of Fe and Bi atoms with respect to the center of the oxgen octahedron. For more simplicity, we can take the relative shift of Fe and Bi atoms from their original positions in the paraelectric phase. Since the space group R3c has an uncertain origin from the crystallographical viewpoint, usually

Bi(z) is fixed as an origin. However, in the present paper we consider the shift of Bi and Fe atoms relative from the oxygen atom, thus we fixed the z-position of the oxygen as an origin (see Fig. 2). In the figure, "s" is the z-component of Bi atom, and then equivalent to the shift of Bi from the cubic position, described as "sc". Meanwhile, "t" is the z-component of Fe atom measured from ¼ position of the cubic phase and the shift of Fe is denoted by "tc". These two shifts of Fe (tc) and Bi (sc) can then be readily translated into the induced electric polarization of the respective elements by multiplying the nominal charges of Fe (3+) and Bi (3+). We note that according to a theoretical calculation [5] the Born effective charges of Fe and Bi are 3.49 and 4.37, respectively. Thus our discussion below can change quantitatively, but not qualitatively.

We have analyzed the newly observed data using a total of over 200 nuclear Bragg peaks. The representative data at two temperatures are given in Fig. 3 (see the Supplemental Material [19] for the summary). We also plotted the temperature dependence of the lattice constants and the rotation angle (ω) of the oxygen octahedron along the c-axis. The lines in Fig 3a and 3b represent our calculated results using the Debye-Gruneissen formula with the same parameters as given in Ref. [16].

The relative shifts of Fe and Bi after being multiplied by the c lattice constant are shown in Fig. 3d as tc and sc. With these measured values of tc and sc, we estimate a total polarization value of ~73 μC/cm$^2$ at room temperature. There are two things noteworthy in the figure. First, tc corresponding to the electric polarization of Fe is as large as half of sc corresponding to the electric polarization of Bi. This implies that both Fe and Bi contribute to the total polarization almost equally. Second, although the Bi shift increases monotonously upon cooling with no hint of an anomaly at $T_N$ there is a clear drop below $T_N$ in the temperature dependence of the Fe shift (tc). This shift of Fe position at $T_N$ bears certain similarity to the Mn off-centering observed in another multiferroic hexagonal manganite at its own antiferromagnetic transition [20]. The decreased electric polarization of Fe clearly then rules out the possibility of the magnetoelectric effect due to the inverse DM effects, leaving a magnetostrictive origin as a probably most likely source for $BiFeO_3$.

In order to analyze the experimental data further, we assume that each of Fe ($P_{Fe}$) and Bi ($P_{Bi}$) electric polarizations follows the usual temperature dependence of 1$^{st}$ order in their respective Ginzburg-Landau (GL) free energies ($F_{Bi}$ and $F_{Fe}$) with the Curie temperature of 1100 K: $F_{Bi/Fe} = \frac{\alpha_2}{2} P_{Bi/Fe}^2 + \frac{\alpha_4}{4} P_{Bi/Fe}^4 + \frac{\alpha_6}{6} P_{Bi/Fe}^6$. In addition, we add a magnetoelectric coupling term of magnetostrictive origin ($F_{ME}$) to the total GL free energy: $F_{ME} = \delta M^2 P_{Fe}^2$. We took the measured temperature-dependent intensity of the (0 0 3) ± Q magnetic superlattice peak as our M in the GL functional: the magnetic superlattice peaks could not be separated because of the instrumental resolution. By using this full GL free energy ($F_{total} = F_{Bi} + F_{Fe} + F_{ME}$) we have tried to fit the temperature dependence of the relative shifts (tc and sc) of Fe and Bi.

First of all, using the GL free energy functional with only $F_{Bi}$ and $F_{Fe}$ terms we have fitted the temperature dependence of both Fe and Bi shifts (tc and sc) above $T_N$ as shown in Fig. 3d. In the case of Bi shift (sc), the experimental data follow well this theoretical line across and even below $T_N$, indicating that the Bi electric polarization ($P_{Bi}$) is not affected by the emergence of the antiferromagnetic order. However, the Fe shift (tc) shows a clear deviation from the theoretical line (dashed line) estimated from above $T_N$: this change in the Fe shift below $T_N$ is denoted as △tc in Fig. 3d. This downwards deviation of tc and so a reduction in the Fe electric polarization ($P_{Fe}$) observed below $T_N$ imply two things of upmost importance. First and foremost, there is an unequivocal coupling between the Fe electric polarization ($P_{Fe}$) and the magnetic moment (M). Second, the Fe electric polarization is suppressed by the onset of the magnetic transition with respect to the values expected of an otherwise nonmagnetic and ferroelectric phase. In order to demonstrate these points, we carried out further analysis now using the full GL free energy functional ($F_{total}$). As shown by the solid line for the Fe shift (tc) below $T_N$, this new approach is clearly successful in explaining the anomaly seen in the temperature dependence of Fe shift (tc).

As a further proof of the existence of the coupling between $P_{Fe}$ and M, we now plot the additional electric polarization of Fe ($\triangle P_{Fe}=q\triangle tc$, where q=+3e for $Fe^{3+}$) against the measured intensity of the (0 0 3)±Q magnetic superlattice peak in Fig. 4. With this new information, we also reanalyzed our previous data taken from another single crystal diffractometer (FONDER), and presented them in Fig. 4 together with the theoretical line as discussed above. As can be seen, both sets of our data from the two different instruments display $\triangle P_{Fe}$, which is in good agreement with one another. It is also consistent with the theoretical calculations of the GL free energy with a negative magnetostrictive manetoelectic coupling, i.e. a positive sign of δ in our Ginzburg-Landau functional. We should note that our results cannot be reconciled by theoretical calculations with an opposite sign of the magnetoelectric coupling as shown by the dashed line in Fig. 4.

We note that our observation of the overall negative magnetoelectric coupling is qualitatively consistent with the field-induced electric polarization observed experimentally [16,21,22]. Notice that all previous pyroelectric current measurements only measured the absolute value of $\triangle P$, not its sign for the intrinsic technical problems. However, we acknowledge there is a discrepancy at a quantitative level. For example, our estimated value of $\triangle P$ is ~ 400 nC/$cm^2$ while the bulk value was reported to be ~ 40 nC/$cm^2$ [16,21]. It may well be plausible that our experimental values of the electric polarization might overestimate the magnetoelectric effect. One passing comment on much wider implications, our works and the conclusion made here suggest that it should be very useful exercises for one to examine the origin of the magnetoelectric coupling by using high-resolution diffraction studies as we have done for $BiFeO_3$ here.

In summary, we have experimentally determined the origin of the magnetoelectric coupling in $BiFeO_3$ by the high-resolution single crystal neutron diffraction studies. Our calculation of the induced electric polarization of each Fe and Bi in the magnetic phase demonstrates that the magnetoelectric coupling of the magnetostrictive origin suppresses the electric polarization at the Fe site below $T_N$, outweighing that of the inverse Dzyaloshinskii-Moriya effects, and becomes the dominant magnetoelectric coupling mechanism for $BiFeO_3$.

**Acknowledgements**
We thank N. Nagaosa, P. Radaelli and R. Johnson for helpful discussions. This work was supported by the Research Center Program of IBS (Institute for Basic Science/Grant No. EM1203) in Korea and by the National Research Foundation of Korea (Grant No. R17-2008-033-01000-0). The work at Rutgers is supported by the US Department of Energy under Contracts No. DOE: DE-FG02-07ER46382.

**References**
[#] jgpark10@snu.ac.kr

[1] M. Fiebig, J. Phys. D **38**, R123 (2005).
[2] W. Eerenstein, N. D. Mathur, and J. F. Scott, Nature **442**, 759 (2006).
[3] J. Wang, J. B. Neaton, H. Zheng, V. Nagarajan, S. B. Ogale, B. Liu, D. Viehland, V. Vaithyanathan, D. G. Schlom, U. V. Waghmare, N. A. Spaldin, K. M. Rabe, M. Wuttig, and R. Ramesh, Science **299**, 1719 (2003).
[4] T. Choi, S. Lee, Y. J. Choi, V. Kiryukhin, and S.-W. Cheong, Science **324**, 63 (2009).
[5] J. B. Neaton, C. Ederer, U. V. Waghmare, N. A. Spaldin, and K. M. Rabe, Phys. Rev. B **71**, 014113 (2005).
[6] I. Sosnowska, T. Peterlin-Neumaier, and E. Steichele, J. Phys. C: Solid State Phys. **15**, 4835 (1982).
[7] J. Jeong, E. A. Goremychkin, T. Guidi, K. Nakjima, Gun Sang Jeon, Shin-Ae Kim, S. Furukawa, Yong Baek Kim, Seongsu Lee, V. Kiryukhin, S-W. Cheong, and Je-Geun Park, Phys. Rev. Lett. **108**, 077202 (2012).
[8] M. Matsuda, R. S. Fishman, T. Hong, C. H. Lee, T. Ushiyama, Y. Yanagisawa, Y. Tomioka, and T. Ito, Phys. Rev. Lett.**109**, 067205 (2012)
[9] K. Ohoyama, S. Lee, S. Yoshii, Y. Narumi, T. Morioka, H. Nojiri, G. S. Jeon, S-W. Jeon, and Je-Geun Park, J. Phys. Soc. Jpn. **80**, 125001 (2011).
[10] S-W. Cheong and M. Mostovoy, Nature Materials **6**, 13 (2007).
[11] H. Katsura, N. Nagaosa, and V. Balatsky, Phys. Rev. Lett. **95**, 057205 (2005).
[12] I. A. Sergienko and E. Dagotto, Phys. Rev. B **73**, 094434 (2006).
[13] J. Rodriguez-Carvajal, Physica B **192**, 55 (1993).
[14] S. Lee, T. Choi, W. Ratcliff II, R. Erwin, S-W. Cheong, and V. Kiryukhin, Phys. Rev. B **78**, 100101(R) (2008).
[15] R. D. Johnson, P. Barone, A. Bombardi, R. J. Bean, S. Picozzi, P. G. Radaelli, Y. S. Oh, S.-W. Cheong, and L. C. Chapon, Phys. Rev. Lett. **110**, 217206 (2013).
[16] J. Park, S-H. Lee, S. Lee, F. Gozzo, H. Kimura, Y. Noda, Y. J. Choi, V. Kiryukhin, S-W. Cheong, Y. Jo, E. S. Choi, L. Balicas, G. S. Jeon, and Je-Geun Park, J. Phys. Soc. Jpn. **80**, 114714 (2011).
[17] S. M. Selbach, T. Tybell, M-A. Einarsrud, and T. Grande, Adv. Mater. **20**, 3692 (2008).
[18] D. C. Arnold, K. S. Knight, F. D. Morrison, and P. Lightfoot, Phys. Rev. Lett. **102**, 027602 (2009).
[19] See Supplemental Materials at http://xxxx for the summary of the crystal structure.
[20] S. Lee, A. Pirogov, M. Kang, K-H. Jang, M. Yonemura, T. Kamiyama, S.-W. Cheong, F. Gozzo, N. Shin, H. Kimura, Y. Noda, and Je-Geun Park, Nature **451**, 805 (2008).
[21] M. Tokunaga, M. Azuma, and Y. Shimakawa, J. Phys. Soc. Jpn. **79**, 064713 (2010).
[22] A. M. Kadomtseva, A. K. Zvezdin, Yu. F. Popov, A. P. Pyatakov, and G. P. Vorob'ev JETP Lett. **79**, 571 (2004).

Figure 1 (Color online) (a) Cycloid magnetic structure of $BiFeO_3$ with the propagation vector Q along the [1 1 0 ] direction. In this hexagonal setting, the total ferroelectric polarization (P) is pointing along the c-axis. (b) and (c) show the temperature dependence of magnetic (1 0 1) and nuclear (0 3 -6) Bragg peaks.

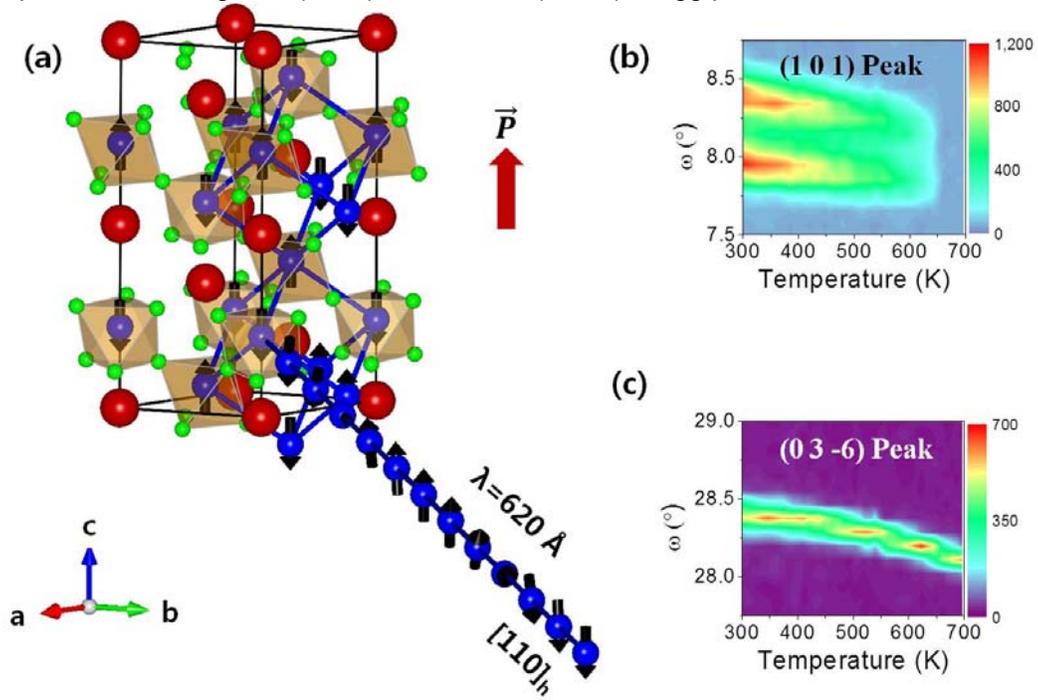

Figure 2 (Color online) (left to right) Paraelectric Pbnm and ideal Pm-3m perovskite structure as well as ferroelectric R3c space group.

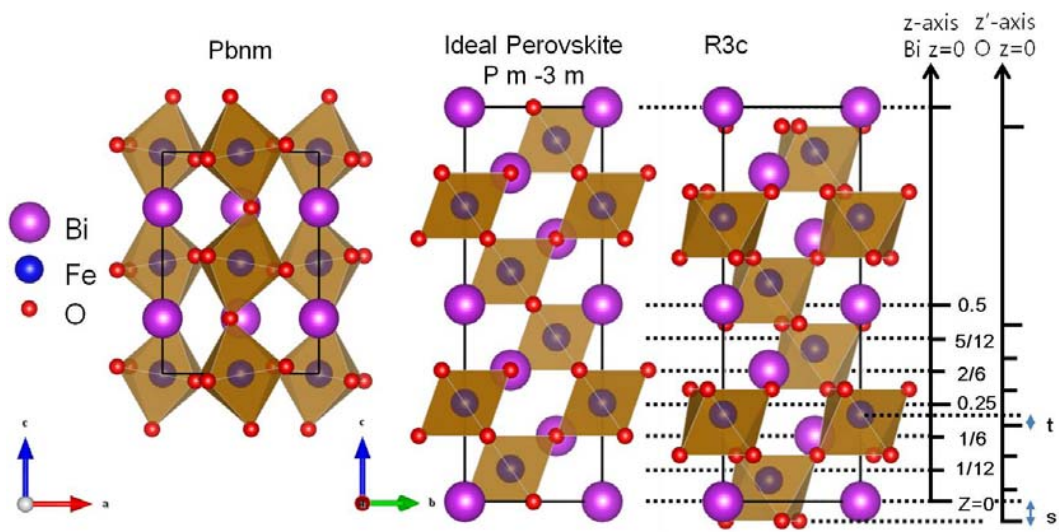

Figure 3 (Color online) (Left) Calculated and observed squared structure factor at two representative temperatures of 300 and 720 K with 222 & 218 nuclear Bragg peaks, respectively. (Right) The temperature dependence is shown of the lattice constants (a and c) together with the antiphase rotation angle ω of the oxygen octahedron in addition to the Bi (sc) and Fe (tc) shifts along the c-axis. The solid lines in (a) and (b) represent our theoretical calculations using a Debye-Gruneissen formula as in the text while the solid lines in (d) represent the theoretical temperature dependence of 1$^{st}$ order ferroelectric order parameter based on the Ginzburg-Landau free energy analysis with the magnetoelectric coupling term as discussed in the text. The dashed line plots the theoretical calculation results of the Fe shift (tc) without the magnetoelectric term.

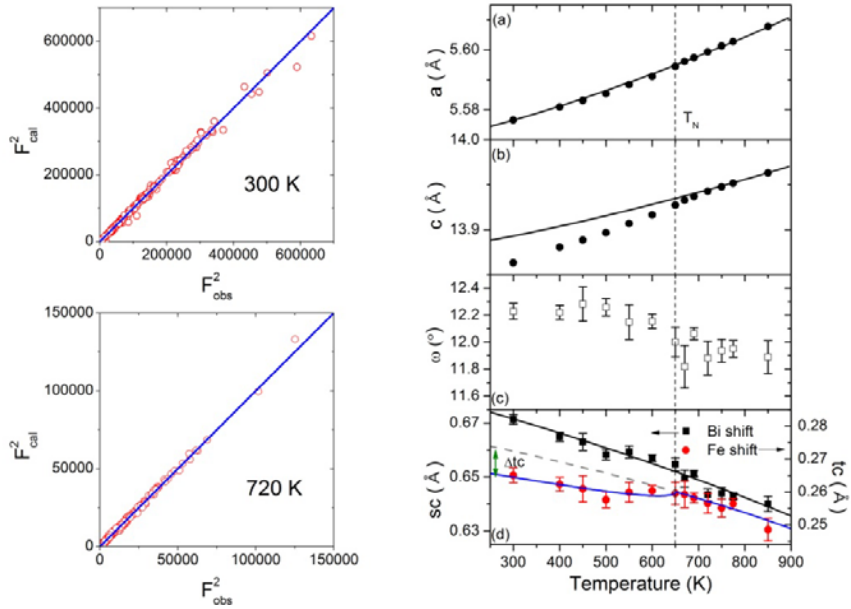

Figure 4 (Color online) Plot of the induced electric polarization of Fe ($\triangle P_{Fe}$) against the measured intensity of the (0 0 3)±Q magnetic superlattice peaks. The solid line represents the theoretical calculation results based on the Ginzburg-Landau free energy analysis with a negative magnetoelectric coupling as discussed in the text while the dashed line does theoretical results expected for the case with the opposite sign for the magnetoelectric coupling.

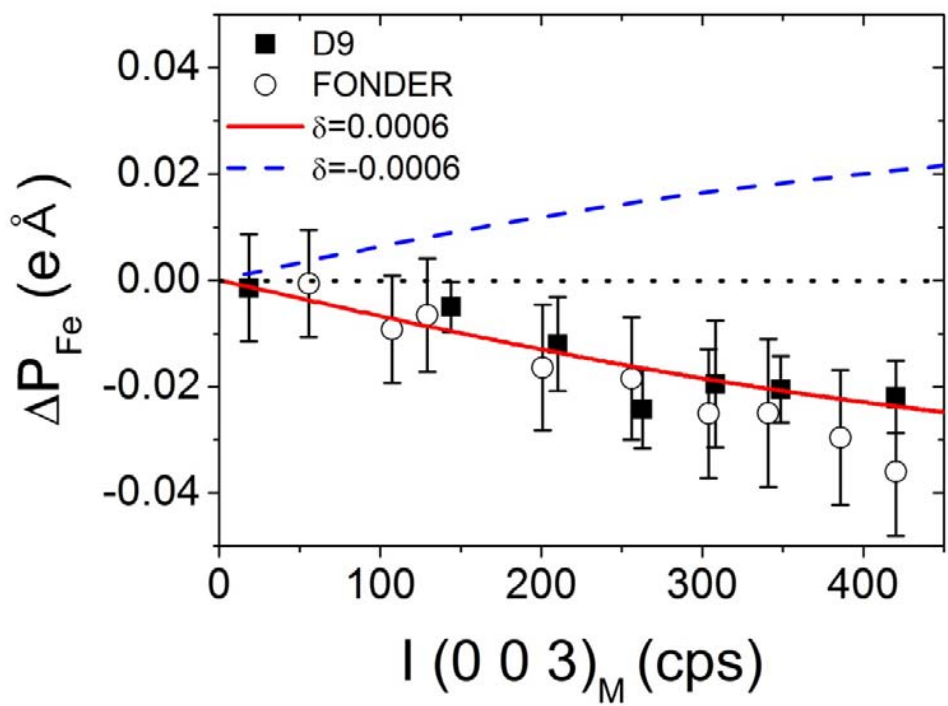